\documentclass[aps,apl,twocolumn,groupedaddress,amsmath]{revtex4}
\usepackage[latin1]{inputenc}    
\usepackage{graphicx}   
\usepackage{xspace}

\begin{document}


\newcommand*{\siX}{\ensuremath{\sigma_\mathrm{x}}\xspace}
\newcommand*{\siZ}{\ensuremath{\sigma_\mathrm{z}}\xspace}

\newcommand*{\PhiX}{\ensuremath{\Phi_\mathrm{X}}\xspace}
\newcommand*{\PhiZ}{\ensuremath{\Phi_\mathrm{Z}}\xspace}
\newcommand*{\fX}{\ensuremath{f_\mathrm{x}}\xspace}
\newcommand*{\fZ}{\ensuremath{f_\mathrm{z}}\xspace}
\newcommand*{\Ax}{\ensuremath{A_\mathrm{x}}\xspace}
\newcommand*{\Az}{\ensuremath{A_\mathrm{z}}\xspace}
\newcommand*{\czx}{\ensuremath{c_\mathrm{zx}}\xspace}

\newcommand*{\TF}{\ensuremath{T_{\varphi \mathrm{F}}}\xspace}
\newcommand*{\TE}{\ensuremath{T_{\varphi \mathrm{E}}}\xspace}
\newcommand*{\GF}{\ensuremath{\Gamma_{\varphi \mathrm{F}}}\xspace}
\newcommand*{\GE}{\ensuremath{\Gamma_{\varphi \mathrm{E}}}\xspace}

\newcommand*{\um}{\ensuremath{\,\mu\mathrm{m}}\xspace}
\newcommand*{\nm}{\ensuremath{\,\mathrm{nm}}\xspace}
\newcommand*{\mm}{\ensuremath{\,\mathrm{mm}}\xspace}
\newcommand*{\m}{\ensuremath{\,\mathrm{m}}\xspace}
\newcommand*{\sqm}{\ensuremath{\,\mathrm{m}^2}\xspace}
\newcommand*{\sqmm}{\ensuremath{\,\mathrm{mm}^2}\xspace}
\newcommand*{\squm}{\ensuremath{\,\mu\mathrm{m}^2}\xspace}
\newcommand*{\psqm}{\ensuremath{\,\mathrm{m}^{-2}}\xspace}
\newcommand*{\psqmV}{\ensuremath{\,\mathrm{m}^{-2}\mathrm{V}^{-1}}\xspace}
\newcommand*{\cm}{\ensuremath{\,\mathrm{cm}}\xspace}

\newcommand*{\nF}{\ensuremath{\,\mathrm{nF}}\xspace}
\newcommand*{\pF}{\ensuremath{\,\mathrm{pF}}\xspace}

\newcommand*{\emob}{\ensuremath{\,\mathrm{m}^2/\mathrm{V}\mathrm{s}}\xspace}
\newcommand*{\edos}{\ensuremath{\,\mu\mathrm{C}/\mathrm{cm}^2}\xspace}
\newcommand*{\mbar}{\ensuremath{\,\mathrm{mbar}}\xspace}

\newcommand*{\A}{\ensuremath{\,\mathrm{A}}\xspace}
\newcommand*{\mA}{\ensuremath{\,\mathrm{mA}}\xspace}
\newcommand*{\nA}{\ensuremath{\,\mathrm{nA}}\xspace}
\newcommand*{\pA}{\ensuremath{\,\mathrm{pA}}\xspace}
\newcommand*{\fA}{\ensuremath{\,\mathrm{fA}}\xspace}
\newcommand*{\uA}{\ensuremath{\,\mu\mathrm{A}}\xspace}

\newcommand*{\Ohm}{\ensuremath{\,\Omega}\xspace}
\newcommand*{\kOhm}{\ensuremath{\,\mathrm{k}\Omega}\xspace}
\newcommand*{\MOhm}{\ensuremath{\,\mathrm{M}\Omega}\xspace}
\newcommand*{\GOhm}{\ensuremath{\,\mathrm{G}\Omega}\xspace}

\newcommand*{\Hz}{\ensuremath{\,\mathrm{Hz}}\xspace}
\newcommand*{\kHz}{\ensuremath{\,\mathrm{kHz}}\xspace}
\newcommand*{\MHz}{\ensuremath{\,\mathrm{MHz}}\xspace}
\newcommand*{\GHz}{\ensuremath{\,\mathrm{GHz}}\xspace}
\newcommand*{\THz}{\ensuremath{\,\mathrm{THz}}\xspace}

\newcommand*{\K}{\ensuremath{\,\mathrm{K}}\xspace}
\newcommand*{\mK}{\ensuremath{\,\mathrm{mK}}\xspace}

\newcommand*{\kV}{\ensuremath{\,\mathrm{kV}}\xspace}
\newcommand*{\V}{\ensuremath{\,\mathrm{V}}\xspace}
\newcommand*{\mV}{\ensuremath{\,\mathrm{mV}}\xspace}
\newcommand*{\uV}{\ensuremath{\,\mu\mathrm{V}}\xspace}
\newcommand*{\nV}{\ensuremath{\,\mathrm{nV}}\xspace}

\newcommand*{\eV}{\ensuremath{\,\mathrm{eV}}\xspace}
\newcommand*{\meV}{\ensuremath{\,\mathrm{meV}}\xspace}
\newcommand*{\ueV}{\ensuremath{\,\mu\mathrm{eV}}\xspace}

\newcommand*{\T}{\ensuremath{\,\mathrm{T}}\xspace}
\newcommand*{\mT}{\ensuremath{\,\mathrm{mT}}\xspace}
\newcommand*{\uT}{\ensuremath{\,\mu\mathrm{T}}\xspace}

\newcommand*{\ms}{\ensuremath{\,\mathrm{ms}}\xspace}
\newcommand*{\s}{\ensuremath{\,\mathrm{s}}\xspace}
\newcommand*{\us}{\ensuremath{\,\mathrm{\mu s}}\xspace}
\newcommand*{\ns}{\ensuremath{\,\mathrm{ns}}\xspace}
\newcommand*{\rpm}{\ensuremath{\,\mathrm{rpm}}\xspace}
\newcommand*{\minute}{\ensuremath{\,\mathrm{min}}\xspace}
\newcommand*{\degree}{\ensuremath{\,^\circ\mathrm{C}}\xspace}

\newcommand*{\EqRef}[1]{Eq.~(\ref{#1})}
\newcommand*{\FigRef}[1]{Fig.~\ref{#1}}
\newcommand*{\dd}[2]{\mathrm{\partial}#1/\mathrm{\partial}#2}
\newcommand*{\ddf}[2]{\frac{\mathrm{\partial}#1}{\mathrm{\partial}#2}}

 \title{Noise correlations in a flux qubit with tunable tunnel coupling}
 \author{Simon Gustavsson$^1$}
 \email{simongus@mit.edu}
 \author{Jonas Bylander$^1$}
 \author{Fei Yan$^2$}
 \author{William D. Oliver$^{1,3}$}
 \affiliation {$^1$Research Laboratory of Electronics, Massachusetts Institute of Technology, Cambridge, MA 02139, USA \\
  $^2$Department of Nuclear Science and Engineering, MIT, Cambridge, MA 02139, USA \\
  $^3$MIT Lincoln Laboratory, 244 Wood Street, Lexington, MA 02420, USA}
 \author{Fumiki Yoshihara$^4$}
 \author{Yasunobu Nakamura$^{4,5}$}
 \affiliation {$^4$The Institute of Physical and Chemical Research (RIKEN), Wako, Saitama 351-0198, Japan \\
 $^5$Green Innovation Research Laboratories, NEC Corporation, Tsukuba, Ibaraki 305-8501, Japan}

\begin{abstract}
We have measured flux-noise correlations in a tunable superconducting flux qubit. The device consists of two loops that  independently control the qubit's energy splitting and tunnel coupling. Low frequency flux noise in the loops causes fluctuations of the qubit frequency and leads to dephasing.  Since the noises in the two loops couple to different terms of the qubit Hamiltonian, a measurement of the dephasing rate at different bias points provides a way to extract both the amplitude and the sign of the noise correlations.  We find that the flux fluctuations in the two loops are anti-correlated, consistent with a model where the flux noise is generated by randomly oriented unpaired spins on the metal surface.
\end{abstract}

\maketitle

The performance of qubits based on superconducting circuits is affected by the ubiquitous low-frequency flux noise present in such devices.
Early measurements on superconducting quantum interference devices (SQUIDs) showed the existence of flux noise with a 
$1/f$-like power spectrum \cite{Koch:1983}.  The noise was found to be universal in the sense that it only depended weakly on materials and sample dimensions \cite{Wellstood:1987}.  
In more recent years, it has become clear that the same excess flux noise limits the phase coherence of superconducting qubits.
A slowly varying flux leads to fluctuations of the qubit energy levels, which has been shown to be the dominant dephasing mechanism in both flux \cite{Yoshihara:2006,Kakuyanagi:2007,Harris:2008} and phase qubits \cite{Bialczak:2007}.  Finding the origin and a possible remedy for the excess flux noise is therefore of  importance to facilitate further improvements of intrinsic coherence times in these devices.

Theories have been developed to explain the microscopic origins of the noise, involving the existence of real or effective unpaired spins in the vicinity of the superconducting structures  \cite{Koch:2007,DeSousa:2007,Faoro:2008,Choi:2009}.
Various experiments have shown the presence of a large number of localized spins on the surface of thin films of normal \cite{Bluhm:2009} and superconducting metals \cite{Sendelbach:2008}, as well as in Si/Si$\mathrm{O}_2$ interfaces \cite{Schenkel:2006}.
Sendelbach \emph{et al.} measured correlations between flux and inductance noise in SQUIDs, possibly suggesting the formation of spin clusters \cite{Sendelbach:2009}. The results suggest that the flux noise is related to the non-equilibrium dynamics of the spin system, possibly described by spin glass models \cite{Chen:2010} or fractal spin clusters \cite{Kechedzhi:2011}.
Recent measurements on qubits with different geometries indicate that the flux noise scales as $(l/w)$, in agreement with the surface spin model \cite{Lanting:2009} ($l$ is the length and $w$ is the width of the superconducting wires). 

\begin{figure}[tb]
\centering
\includegraphics[width=\linewidth]{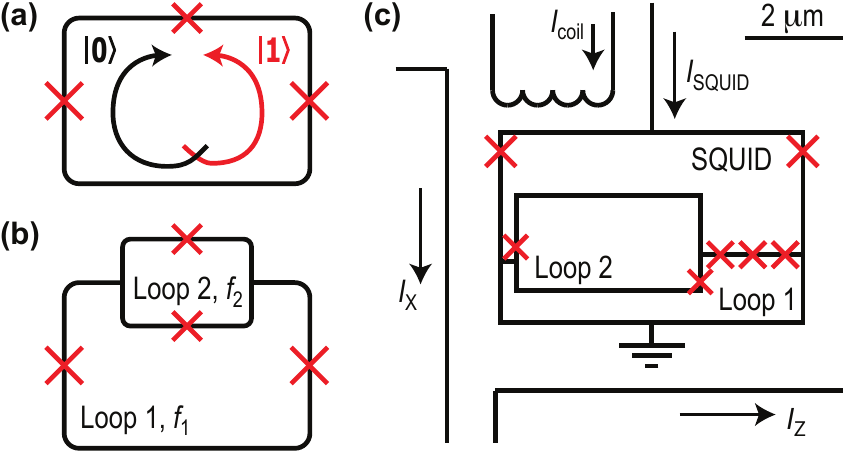}
\caption{(a) A standard flux qubit.  The two diabatic states correspond to clockwise and counter-clockwise circulating currents.
 (b) A tunable flux qubit. The smallest junction has been replaced by an additional loop.
 (c) The actual sample design. The qubit is embedded in a SQUID used for read-out of the qubit state. The line widths are not drawn to scale. 
} \label{fig:Sample}
\end{figure}

Flux noise correlations have been studied in a system of coupled qubits sharing parts of their loops \cite{Yoshihara:2010}.
It was found that the flux fluctuations originating from the shared branch lead to correlations in the noise of the two qubits.
In this work, we use a single, two-loop qubit to investigate flux noise correlations between different parts within a single qubit [\FigRef{fig:Sample}(b)].
The fluxes in the two loops couple to the longitudinal and transverse components of the qubit's Hamiltonian, respectively.
The ability to control both parameters of the Hamiltonian allows us to investigate qubit coherence properties at different frequencies while remaining at the optimal bias point \cite{Paauw:2009}.
By measuring the qubit dephasing rate as a function of flux bias, we can extract the correlations between the flux fluctuations in the two loops.
Knowing that the two fluxes couple differently to the qubit energy allows us to determine the sign of the correlations.
We find the flux noise to be strongly anti-correlated, in agreement with a model where the noise is generated by spins on the superconductor surface.

The standard flux qubit consists of a superconducting loop with three or more Josephson junctions [\FigRef{fig:Sample}(a)]. The diabatic states correspond to clockwise and counterclockwise circulating currents, respectively \cite{Mooij:1999,Orlando:1999}. When the flux in the loop is close to half a flux quantum, $\Phi = 0.5  \Phi_0$, the Hamiltonian is $H = - (\frac{\Delta}{2}\siX + \frac{\varepsilon}{2}\siZ)$ within a two-level approximation. Here, $\varepsilon = 2 I_\mathrm{P} (\Phi-0.5\Phi_0)$ where $I_\mathrm{P}$ is the persistent current and $\Phi_0= h/2e$.  The tunnel coupling $\Delta$ is fixed by fabrication and determined by the size of the smallest Josephson junction.
By replacing the smallest junction with a second loop, the tunnel coupling $\Delta$ can be tuned \emph{in situ} by the flux $\Phi_2 = f_2 \Phi_0$ in the second loop [\FigRef{fig:Sample}(b)] \cite{Paauw:2009,Zhu:2010}.
Due to the device geometry, the detuning $\varepsilon$ is controlled by the combined flux $(f_1 + f_2/2)\Phi_0$, where $f_1 = \Phi_1/\Phi_0$ is the normalized flux in loop 1 \cite{Orlando:1999}. To clarify the relation between the qubit parameters and the two fluxes, we introduce the effective fluxes
\begin{equation}  \label{eq:Fluxes}
 \fZ = (f_1 + f_2/2) -1/2 ,~~~~~~~ \fX = f_2.
\end{equation}
The energy separation of the qubit is then given by
\begin{equation}  \label{eq:Energy}
 E_{01}/h = \sqrt{\varepsilon^2 + \Delta^2}, \mathrm{~with~} \varepsilon = \varepsilon(\fZ),~\Delta = \Delta(\fX).
\end{equation}
The qubit is embedded in a dc SQUID for reading out the qubit state.  The read-out is implemented by applying a short current pulse to the SQUID; due to the inductive coupling between the SQUID and the qubit, the SQUID switching probability will vary depending on the qubit state \cite{Chiorescu:2003}.

The design of the actual device is shown in \FigRef{fig:Sample}(c) (see the supplementary material for a detailed drawing of the device \cite{suppMatDrawing}). The structure is made of aluminum and is fabricated using e-beam lithography and shadow evaporation.   There are two local current-bias lines, which, together with the global external field, allows us to control the fluxes in the two qubit loops and in the read-out SQUID independently. 
Due to the close proximity of the loops and the bias lines there is substantial cross-coupling between the different elements. During the measurements we applied appropriate compensation currents to the bias lines to compensate the unwanted couplings. The residual unwanted coupling is less than $1\%$.

\begin{figure}[tb]
\centering
\includegraphics[width=\linewidth]{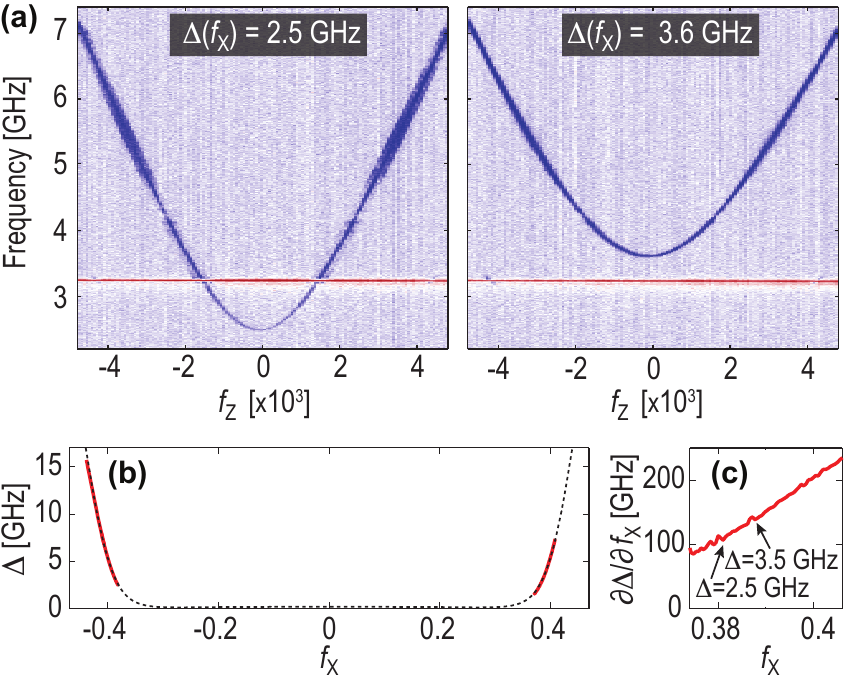}
\caption{(a) Qubit spectra, measured for two different values of $\fX$. The features at 3.2 GHz are due to the SQUID plasma frequency.
 (b) Qubit tunnel coupling $\Delta$ measured as a function of $\fX$. The dashed line is the result of a numerical simulation.
 (c) Numerical derivative $\dd{\Delta}{\fX}$ of the data in (b). 
} \label{fig:TuneDelta}
\end{figure}

Figure \ref{fig:TuneDelta}(a) shows two qubit spectra measured for two values of $\fX$. The measurement was done by applying a long microwave pulse ($3\us$) to saturate the qubit before reading out its state.
The spectra have the form expected from \EqRef{eq:Energy}, but with different values of $\Delta$. The horizontal feature around 3.2 GHz is due to the plasma mode of the read-out SQUID. In \FigRef{fig:TuneDelta}(b) we plot the measured $\Delta$ as a function of $\fX$.
The tunnel coupling $\Delta$ is symmetric around $\fX=0$ and can be tuned experimentally from $1.5$ to $15\GHz$. The dashed line shows the results of a numerical simulation of the qubit energy levels for realistic fabrication parameters. To quantify the relation between $\Delta$ and $\fX$ we plot the numerical derivative of the data for a region around $\fX = 0.39$ [\FigRef{fig:TuneDelta}(c)].  We have marked the positions of the two operating points with $\Delta = 2.5\GHz$ and $\Delta = 3.5\GHz$ where we will perform coherence measurements. The sensitivity $\dd{\Delta}{\fX}$ is different for those two points.

\begin{figure}[tb]
\centering
\includegraphics[width=\linewidth]{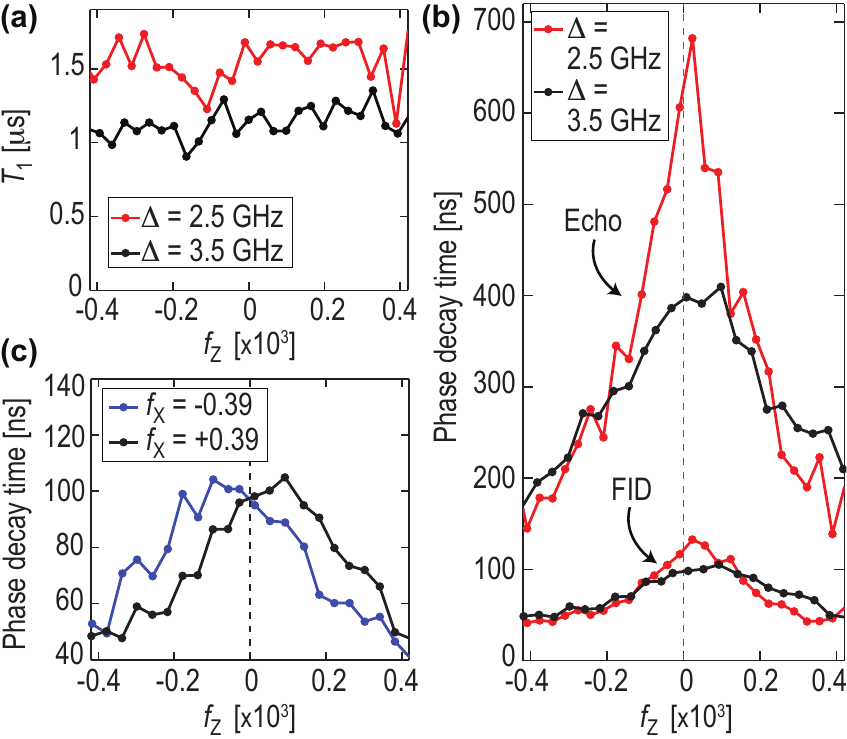}
\caption{
 (a) Energy decay time $\mathrm{T_1}$.  The decay does not depend strongly on the flux detuning.
 (b) Echo and free-induction decay times versus flux detuning.  The position of the longest decay times is shifted to positive flux detuning.
 (c) Free-induction decay measured at  $\fX = \pm 0.39$.  Depending on the sign of $\fX$, the position of the longest decay time shifts in different directions. All measurements were done by sweeping $\fZ$ while keeping $\fX$ constant.
} \label{fig:T1T2}
\end{figure}

The results of \FigRef{fig:TuneDelta} show that our device behaves as expected and that we can tune the qubit parameters $\Delta$ and $\varepsilon$ independently by applying fluxes in the two loops. To characterize the qubit coherence as a function of those parameters, we perform energy-relaxation ($T_1$), free-induction decay (FID) and Hahn-echo measurements.  
$T_1$ is measured by applying a $\pi$ pulse to the qubit and delaying the read-out.  Figure \ref{fig:T1T2}(a) shows the extracted $T_1$ decay times as a function of detuning $\fZ$, measured for the two values of $\Delta$ marked in \FigRef{fig:TuneDelta}(c).  $T_1$ is fairly independent of $\fZ$.
The difference in $T_1$ for the two values of $\Delta$ can be attributed to differences in detuning from the SQUID plasma mode at $f_\mathrm{plasma} = 3.2\GHz.$

The FID sequence consists of two $\pi/2$ pulses separated by a time $t$. The FID is sensitive to low-frequency noise that causes adiabatic fluctuations in the qubit energy splitting $E_{01}$. The Hahn-echo contains an extra $\pi$ pulse in the middle of the sequence to refocus low-frequency fluctuations.
For Gaussian-distributed noise with a $1/f$-type spectrum, the decay of both the FID and the Hahn-echo has the form $p(t) = e^{-t/2T_1} \, e^{-(t/T_{\varphi \mathrm{(F/E)}})^2}$. By first measuring  $T_1$ we can extract the pure dephasing times $\TF$ and $\TE$ \cite{Yoshihara:2006}.  

In \FigRef{fig:T1T2}(b) we plot $\TF$ and $\TE$ as a function of $\fZ$ for the two values of $\Delta$. We make a few observations:
(i) The time $\TE$ is 4-5 times longer than $\TF$, consistent with the $1/f$-type spectrum  \cite{Ithier:2005}. 
(ii) Both $\TF$ and $\TE$ fall off as $|\fZ|$ is increased; this is because 
$\dd{E_{01}}{\fZ}$ is zero only at $\fZ=0$ and increases approximately linearly with $|\fZ|$ over this range of $\fZ$.  However, contrary to Refs.~\cite{Yoshihara:2006, Bylander:2011}, the echo dephasing times are considerably shorter than the limit set by energy relaxation ($2T_1$) even close to $\fZ=0$.  This is a result of our device having an extra loop that controls $\Delta$: fluctuations in $\fX$ will couple to $\Delta$, which will couple to $E_{01}$ even when $\fZ=0$.
(iii) Close to $\fZ=0$, the dephasing times are longer for $\Delta = 2.5\GHz$ than for $\Delta = 3.5\GHz$: this is because the sensitivity $\dd{\Delta}{\fX}$ to noise in $\fX$ is stronger at $\Delta = 3.5\GHz$ [see \FigRef{fig:TuneDelta}(c)].
(iv) The longest dephasing times do not occur at $\fZ=0$ but are slightly shifted to positive $\fZ$.  The effect is visible for both values of $\Delta$, but the shift is larger for $\Delta = 3.5\GHz$.  

We investigate the shift in more detail by measuring the FID at $\Delta = 3.5\GHz$ for positive and negative values of $\fX$. The results are shown in \FigRef{fig:T1T2}(c); at  $\fX = -0.39$, the longest dephasing times are shifted towards \emph{negative} $\fZ$. Even though $\Delta$ is the same at both $\fX = 0.39$ and $\fX = -0.39$, the sensitivity $\dd{\Delta}{\fX}$ is negative for $\fX<0$ and positive for $\fX>0$.  The fact that the shift depends on the sign of $\dd{\Delta}{\fX}$ leads us suspect that there are correlations between $\fX$ and $\fZ$.  We have checked that the shift does not depend on SQUID bias current for small excursions from the operating point \cite{Bertet:2005}.

\begin{figure}[tb]
\centering
\includegraphics[width=\linewidth]{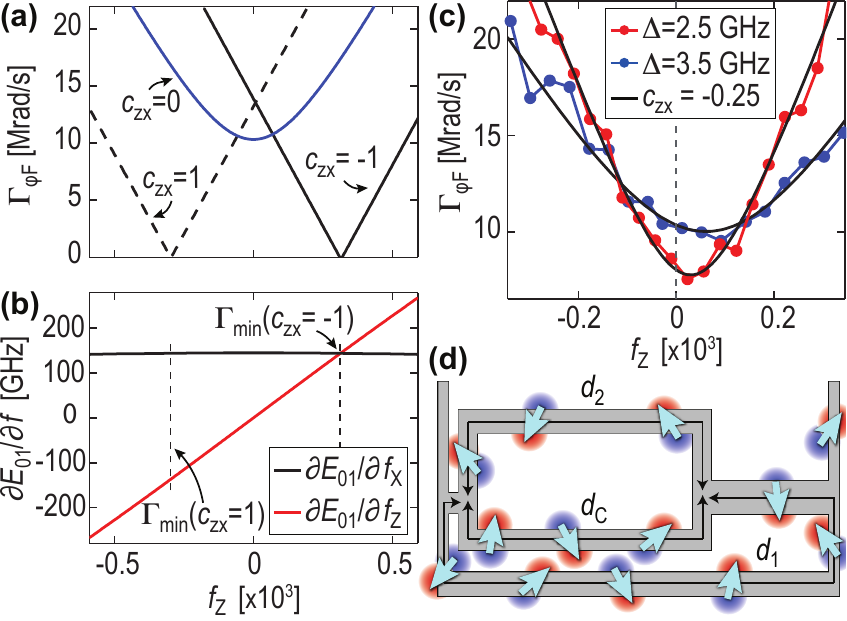}
\caption{
 (a) Phase decay rate $\GF$ expected from the sensitivities shown in (b), with $\Az = \Ax = (3\,\mathrm{\mu\Phi_0^2})$. 
 (b) Sensitivity of the qubit energy to flux fluctuations. For perfectly correlated noise ($\czx = \pm 1$), the fluctuations cancel out when $\dd{E_{01}}{\fX} = \mp \dd{E_{01}}{\fZ}$.
 (c) Measured decay rate $\GF$.  The fit gives a correlation factor of $\czx = -0.25$.
 (d) Schematic picture of the surface-spin model.  Spins located on the shared line marked by $d_c$ induce fields pointing in opposite direction in the two loops. 
} \label{fig:NoiseXZ}
\end{figure}

The phase decay rate $\GF = 1/\TF$ is due to a combination of noise in both $\fX$ and $\fZ$.  To calculate $\GF$ in the presence of correlations, we assume that the fluctuations $\delta\!\fX$ and $\delta\!\fZ$ are described by noise spectra of the form $S_{\fX }(\omega)= (\Ax\Phi_0^{-2})/|\omega|$ and $S_{\fZ }(\omega) = (\Ax\Phi_0^{-2})/|\omega|$. In addition we introduce the correlation spectrum $S_{\fZ\fX}(\omega) = (A_\mathrm{zx}\Phi_0^{-2})/|\omega|$ \cite{Hu:2007, Yoshihara:2010} and the correlation coefficient $\czx =  \langle \delta\!\fZ \, \delta\!\fX \rangle/\sqrt{\langle \delta\!\fZ^2\rangle \, \langle \delta\!\fX^2\rangle} = A_\mathrm{zx}/\sqrt{\Ax \Az}$.  The total decay rate is \cite{Ithier:2005}
\begin{eqnarray}  \label{eq:TotalDecay}
  \GF &=& \frac{\ln(1/\omega_\mathrm{low} t)}{\hbar \, \Phi_0^{2}} \bigg[  \Ax \left( \ddf{E_{01}}{\fX}\right)^2 + 
  \Az \left( \ddf{E_{01}}{\fZ}\right)^2 +  \nonumber \\
 && 2\, \czx\, \sqrt{\Ax \Az} \left( \ddf{E_{01}}{\fX} \right) \left( \ddf{E_{01}}{\fZ} \right) \bigg]^{1/2}.
\end{eqnarray}
The low-frequency cut-off $\omega_\mathrm{low}/2\pi = 1\Hz$ is imposed by the measurement protocol. 
Figure~\ref{fig:NoiseXZ}(a) shows the result of \EqRef{eq:TotalDecay}, plotted for $\Az = \Ax$ and three different values of the correlation coefficient $\czx$. Without any correlations ($\czx\!=\!0$), the decay rate increases with $|\fZ|$, and the value at $\fZ=0$ is set by the noise in $\fX$.
However, if we assume perfectly correlated ($\czx\!=\!1$) or anti-correlated noise ($\czx\!=\!-1$), the point of minimal decay shifts to negative and positive values of $\fZ$, respectively. In addition, the decay rate goes to zero in those points.  

To understand how correlations can lead to shifts of the dephasing times, we consider how the fluctuations in $\fX$ and $\fZ$ couple to the energy splitting $E_{01}$.
Figure~\ref{fig:NoiseXZ}(b) shows the sensitivities $\dd{E_{01}}{\fX}$ and $\dd{E_{01}}{\fZ}$ for $\Delta = 3.5\GHz$, extracted from the data in \FigRef{fig:TuneDelta}.  If the noise is completely correlated, the  fluctuations in $E_{01}$ will perfectly cancel when $\dd{E_{01}}{\fX} = - \dd{E_{01}}{\fZ}$. This happens at the point marked by $\Gamma_\mathrm{min}(\czx\!=\!1)$ in \FigRef{fig:NoiseXZ}(b). On the other hand, if the noise is perfectly anti-correlated, the  fluctuations in $E_{01}$ will cancel when $\dd{E_{01}}{\fX} = \dd{E_{01}}{\fZ}$.
This also explains why the phase decay data taken at $\fX = -0.39$ shifts in the opposite direction [\FigRef{fig:T1T2}(c)].  Here, the derivative $\dd{\Delta}{\fX}$ is negative, which gives $\dd{E_{01}}{\fX}$ a minus sign and makes the points $\Gamma_\mathrm{min}(\czx\!=\!1)$ and $\Gamma_\mathrm{min}(\czx\!=\!-1)$ trade places.

We can fit the measured decay rate to \EqRef{eq:TotalDecay} in order to extract the flux-noise amplitude and the correlation coefficient in our sample.  The results are shown in \FigRef{fig:NoiseXZ}(c), giving $\Az = (3.9\,\mathrm{\mu\Phi_0)^2}$, $\Ax = (3.1\,\mathrm{\mu\Phi_0)^2}$  and $c = -0.25 \pm 0.05$. We stress that the same fitting parameters are used for both values of $\Delta$.  The reason why the data for $\Delta=2.5\GHz$ shows a less dramatic shift in $\fZ$ is because $\dd{E_{01}}{\fX}$ is smaller at this value of $\Delta$.
The values for $\dd{E_{01}}{\fX}$  and $\dd{E_{01}}{\fZ}$ used in the fits were extracted from \FigRef{fig:TuneDelta}, giving $\dd{\varepsilon}{\fZ} = 1.37\THz$, $\dd{\Delta}{\fX} = 110\GHz$ for $\Delta = 2.5\GHz$ and $\dd{\varepsilon}{\fZ} = 1.28\THz$, $\dd{\Delta}{\fX} = 142\GHz$ for $\Delta = 3.5\GHz$. 

It is important to point out that the correlations discussed so far have been between fluctuations $\delta\!\fZ$, $\delta\!\fX$ in the \emph{effective} fluxes defined by \EqRef{eq:Fluxes}.  The fluctuations and the correlations in the \emph{geometric} fluxes $f_1$, $f_2$ can be found by inverting \EqRef{eq:Fluxes}.  Assuming the fluctuations $\delta\!f_1$, $\delta\!f_2$ are described by spectra $S_{f_1}(\omega) = (A_{1}\Phi_0^{-2})/|\omega|$ and $S_{f_2}(\omega) = (A_{2}\Phi_0^{-2})/|\omega|$, we find \cite{suppMat}
\begin{eqnarray}  \label{eq:ConvCorr}
  A_1  & = & \Az + \Ax/4 - \czx \sqrt{\Az \Ax} = (4.5\,\mathrm{\mu\Phi_0)^2} \nonumber \\ 
  A_2 & = & \Ax = (3.1\,\mathrm{\mu\Phi_0)^2}, \nonumber \\ 
  c_{12} &=& 
  \czx \sqrt{ \frac{ \Az}{A_1}} - \frac{1}{2} \sqrt{ \frac{\Ax}{A_1}} = -0.55. 
\end{eqnarray}
Fabrication imperfections may cause deviations in the sizes of the two junctions in loop 2, which will affect \EqRef{eq:ConvCorr}.  From the results of \FigRef{fig:TuneDelta}(b), we know that the junction asymmetry is less than $\pm10\%$, which will lead to an uncertainty of the correlation coefficient $c_{12}$ of about $\pm0.1$.  Note that the measurement is sensitive to correlations in the frequency range $1\Hz$ to 
$\sim\!1\MHz$, as set by the FID weighting function \cite{Ithier:2005,Bylander:2011}.

The results of \EqRef{eq:ConvCorr} show that the flux fluctuations in the two loops of the sample are anti-correlated.
This is clear evidence that the noise is not due to a global fluctuating magnetic field, since such fluctuations would give positive correlations.
Instead, we consider a model where the noise is due to a large number of randomly oriented spins distributed over the  metal surface [see \FigRef{fig:NoiseXZ}(d)]. The spins will generate local magnetic fields that is picked up by the loops, and fluctuations of the spin orientations will give rise to flux noise \cite{Koch:2007}.
In our sample, spins that are located on the line that is shared by the loops [marked by $d_\mathrm{c}$ in \FigRef{fig:NoiseXZ}(d)] will induce fields of opposite directions in loop 1 and loop 2. This will give rise to negative correlations between the fluxes in the loops.

To get an estimate of the expected correlations, we assume (i) that the field of a single spin only couples to a loop if it is sitting on that loop's circumference, (ii) that the width of the sample wires is constant, and (iii) that the amount of flux that a loop picks up from a single spin is independent of the spin's position along the width of the wire.  The assumptions seem reasonable based on numerical simulations of the coupling between a SQUID and the magnetic moment of a single spin \cite{Koch:2007}.  Under these conditions the flux fluctuations generated by an ensemble of randomly oriented spins will scale as $\delta\!f \propto \sqrt{\rho \, d}$, where $\rho$ is the spin density and $d$ the length of the loop. The correlation coefficient becomes
\begin{equation}  \label{eq:CorrModel}
  c_{12} = \frac{\langle \delta\!f_1 \, \delta\!f_2 \rangle}{\sqrt{\langle \delta\!f_1^2\rangle \, \langle \delta\!f_2^2\rangle}} =
  \frac{-d_c}{\sqrt{(d_1+d_c)(d_2+d_c)}}.
\end{equation}
Here $d_\mathrm{c}$ is the length of the line segment shared by both loops, and $d_1$, $d_2$ are the lengths of remaining sections of the two loops [\FigRef{fig:NoiseXZ}(d)].  With values relevant for our sample $(d_1 = 10.8 \um, ~d_2 = 6.0 \um, ~d_c = 4.8 \um)$, we find $c_{12} = -0.37$.  This is smaller than the value extracted experimentally, but the agreement is reasonable considering the simplicity of the model.  Further work is needed to develop a more realistic model of the device \cite{suppMatDrawing}, and to describe the correlations when spin interactions are taken into account \cite{Chen:2010,Kechedzhi:2011}.

To conclude, we have investigated flux noise correlations by measuring dephasing rates in a tunable flux qubit.  We find the flux fluctuations in neighboring loops to be anti-correlated, in agreement with predictions from models where the flux noise is generated by randomly oriented spins on the metal surface. 
The ability to extract noise correlations provides important information about the microscopic origin of the flux noise.  The method presented here is general and can be extended to samples with different geometries.  Combining this method with recent improvements in multi-pulse noise spectroscopy \cite{Bylander:2011} opens up the possibility to measure correlations at different frequencies and thereby probe the dynamics of the surface spin system.%

We thank T. Orlando and R. McDermott for helpful discussions, K. Harrabi for assistance with device fabrication and P. Forn-D\'{\i}az for performing numerical simulations of the qubit spectrum.

\bibliographystyle{apsrev}
\bibliography{TunableQubit}

\clearpage

\section*{Supplementary material}
\subsection*{Device drawing}
\begin{figure}[h]
\centering
\includegraphics[width=\linewidth]{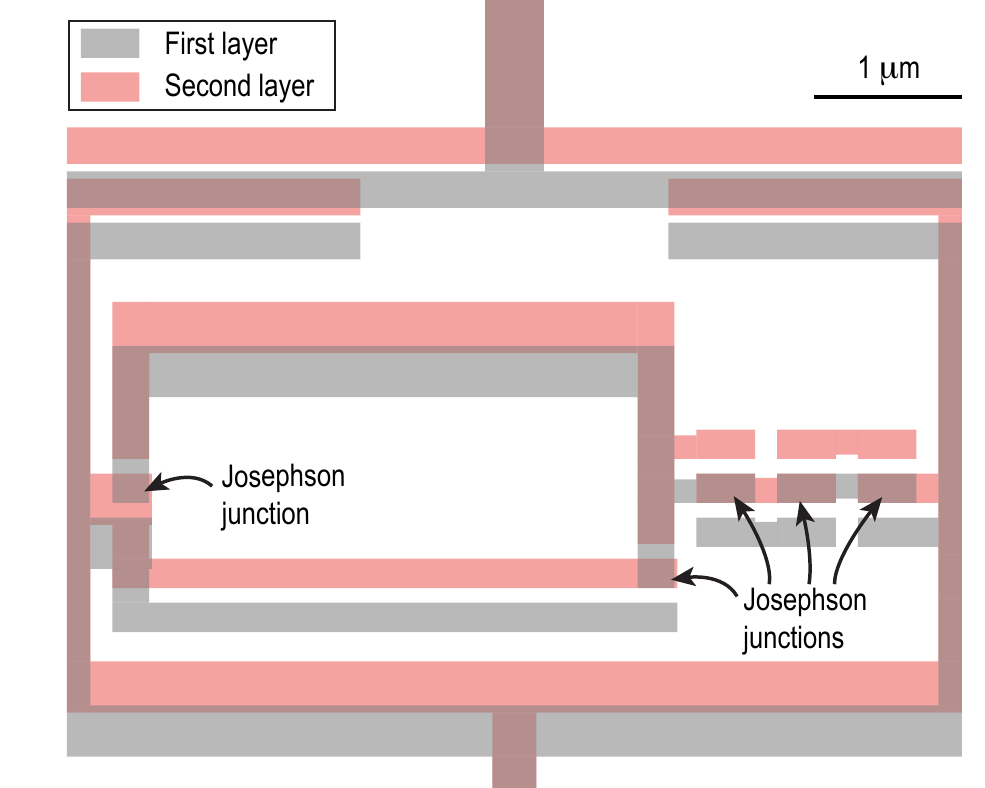}
\caption{Drawing of the device, showing the location of metal from the two evaporation stages.
} \label{fig:CAD}
\end{figure}

The device was fabricated using standard two-angle shadow evaporation techniques. The two layers of metal are deposited from different angles, resulting in overlapping superconductors separated by a thin film of insulating aluminum oxide.  The method is used to form the Josephson junctions in the system (see \FigRef{fig:CAD}). 

\subsection*{SEM picture of the device}
\begin{figure}[h]
\centering
\includegraphics[width=\linewidth]{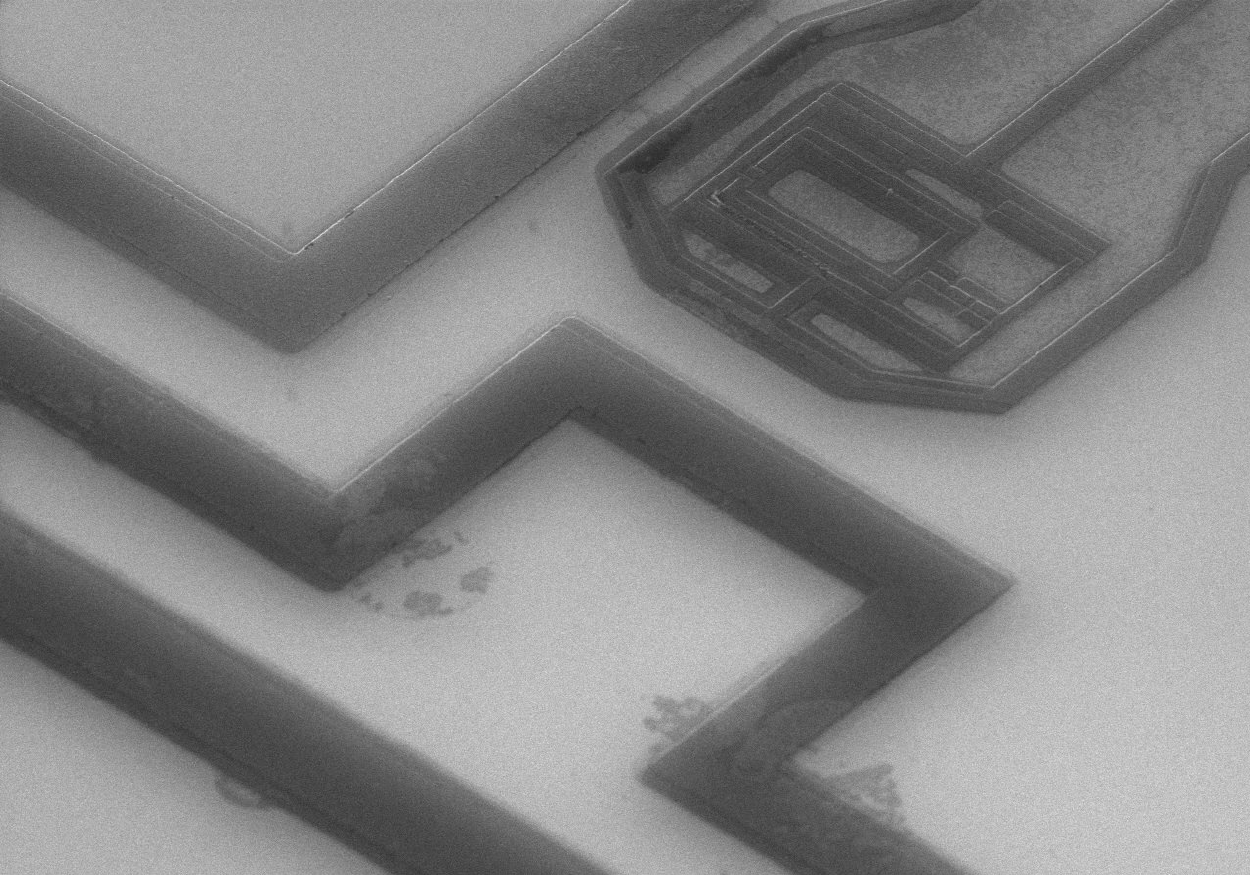}
\caption{SEM picture of a device with the same geometry as the one used in the experiments.
} \label{fig:SEM}
\end{figure}

Figure \ref{fig:SEM} shows a SEM picture of a device with the same geometry as the one used in the experiments.
The qubit and the read-out SQUID are in the upper-right part of the figure.  The two lines next to the qubit are the local flux bias lines, while the thicker line in the lower-left part of the figure is the microwave drive line.

\subsection*{Derivation of flux correlations}
We start by defining the normalized fluxes $f_1 = \Phi_1/\Phi_0$ and $f_2 = \Phi_2/\Phi_0$ in the two qubit loops.   
Correlations between fluctuations $\delta\!f_1$ and $\delta\!f_2$ in those parameters are characterized by the correlation coefficient $c_{12}$, defined as
\begin{equation} \label{eq:c12}
 c_{12} = \frac{\langle \delta\!f_1 \, \delta\!f_2 \rangle} 
 {\sqrt{\langle  \delta\!f_1^2 \rangle \, \langle  \delta\!f_2^2 \rangle}}.
\end{equation}
The qubit parameters $\varepsilon = \varepsilon(\fZ)$ and $\Delta = \Delta (\fX)$ are controlled by the effective fluxes
\begin{eqnarray} \label{eq:defFz}
 \fZ &=& (f_1 + f_2/2) -1/2,  \nonumber \\
 \fX &=& f_2.
\end{eqnarray}
The coefficient for describing correlations between the fluctuations $\delta\!\fZ$ and $\delta\!\fX$ in the effective fluxes is given by
\begin{eqnarray} \label{eq:cZX}
 \czx &=& \frac{\langle \delta\!\fZ \, \delta\!\fX \rangle} 
 {\sqrt{\langle  \delta\!\fZ^2 \rangle \, \langle  \delta\!\fX^2 \rangle}} = 
 \frac{\langle (\delta \! f_1 + \delta\!f_2/2)  \, \delta\! f_2 \rangle} 
 {\sqrt{\langle  \delta\!\fZ^2 \rangle \, \langle  \delta\!f_2^2 \rangle}} = \nonumber \\
&=&  \frac{\langle \delta \! f_1 \, \delta\! f_2 \rangle}  
 {\sqrt{\langle  \delta\!\fZ^2 \rangle \, \langle  \delta\!f_2^2 \rangle}} + 
 \frac{\langle \delta \! f_2^2 \rangle /2}  
 {\sqrt{\langle  \delta\!\fZ^2 \rangle \, \langle  \delta\!f_2^2 \rangle}}.
\end{eqnarray}
We want to be able to extract $c_{12}$ from the measured correlations $\czx$ between the effective fluxes. We start by inserting the expression for $\langle \delta\!f_1 \, \delta\!f_2 \rangle$ from \EqRef{eq:c12} into \EqRef{eq:cZX}.  This gives
\begin{eqnarray} \label{eq:cZX2}
 \czx &=& \frac{c_{12} \sqrt{\langle  \delta\!f_1^2 \rangle \, \langle  \delta\!f_2^2 \rangle} }  
 {\sqrt{\langle  \delta\!\fZ^2 \rangle \, \langle  \delta\!f_2^2 \rangle}} + 
 \frac{\langle \delta \! f_2^2 \rangle /2}  
 {\sqrt{\langle  \delta\!\fZ^2 \rangle \, \langle  \delta\!f_2^2 \rangle}} = \nonumber \\
 &=& c_{12} \sqrt{ \frac{ \langle  \delta\!f_1^2 \rangle} {\langle  \delta\!\fZ^2 \rangle }} + 
 \frac{1}{2} \sqrt{ \frac{\langle \delta \! f_2^2 \rangle }  
  {\langle  \delta\!\fZ^2 \rangle}}.
\end{eqnarray}
Solving for $c_{12}$ gives
\begin{equation} \label{eq:c12_cZX}
 c_{12} = \czx \sqrt{ \frac{ \langle  \delta\!\fZ^2 \rangle} {\langle  \delta\!f_1^2 \rangle }} - 
 \frac{1}{2} \sqrt{ \frac{\langle \delta \! f_2^2 \rangle }  {\langle  \delta\! f_1^2 \rangle}} = 
 \czx \sqrt{ \frac{ \langle  \delta\!\fZ^2 \rangle} {\langle  \delta\!f_1^2 \rangle }} - 
 \frac{1}{2} \sqrt{ \frac{\langle \delta \! \fX^2 \rangle }  {\langle  \delta\! f_1^2 \rangle}}.
\end{equation}
In addition, we need to express the fluctuations $\langle  \delta\! f_1^2 \rangle$ in terms of the effective fluctuations $\czx$, $\langle  \delta\! \fX^2 \rangle$ and $\langle  \delta\! \fZ^2 \rangle$.  From the definition in \EqRef{eq:defFz}, we have
\begin{equation} \label{eq:avfZ}
  \langle  \delta\!\fZ^2 \rangle = \langle (\delta \! f_1 + \delta\!f_2/2)^2 \rangle = 
  \langle \delta \! f_1^2 \rangle + \langle \delta \! f_1 \, \delta \! f_2 \rangle + \langle \delta \! f_2^2 \rangle/4.
\end{equation}
Using the expression for $\langle \delta\!f_1 \, \delta\!f_2 \rangle$ from \EqRef{eq:c12}, we get
\begin{equation} \label{eq:avfZ2}
  \langle  \delta\!\fZ^2 \rangle = 
  \langle \delta \! f_1^2 \rangle + c_{12} \sqrt{\langle  \delta\!f_1^2 \rangle \, \langle  \delta\!f_2^2 \rangle} + \langle \delta \! f_2^2 \rangle/4.
\end{equation}
Inserting the expression for $c_{12}$ from \EqRef{eq:c12_cZX} gives
\begin{equation} \label{eq:avfZ2}
  \langle  \delta\!\fZ^2 \rangle = 
  \langle \delta \! f_1^2 \rangle + 
  \czx \sqrt{\langle \delta\!\fZ^2 \rangle \langle \delta\!f_2^2 \rangle} -  
  \frac{1}{2} \sqrt{\langle \delta\!\fX^2 \rangle \langle \delta\!f_2^2 \rangle}
  + \langle \delta \! f_2^2 \rangle/4.
\end{equation}
Noting that $\langle \delta\! f_2^2 \rangle = \langle  \delta\!\fX^2 \rangle$, we have
\begin{equation} \label{eq:avfZ3}
  \langle  \delta\!\fZ^2 \rangle = 
  \langle \delta \! f_1^2 \rangle + 
  \czx \sqrt{\langle \delta\!\fZ^2 \rangle \langle \delta\! \fX^2 \rangle} -  
  \langle \delta \! \fX^2 \rangle/4.
\end{equation}
Finally, solving for $\langle  \delta\! f_1^2 \rangle$ gives
\begin{equation} \label{eq:avf1}
  \langle  \delta\! f_1^2 \rangle = 
  \langle \delta \! \fZ^2 \rangle + \langle \delta \! \fX^2 \rangle/4 -
  \czx \sqrt{\langle \delta\!\fZ^2 \rangle \langle \delta\! \fX^2 \rangle}.
\end{equation}
We assume that the fluctuations $\delta\!f_1$, $\delta\!f_2$, $\delta\!\fZ$ and $\delta\!\fX$ are all described by noise spectra of the form $S_{f_1}(\omega)= A_1/|\omega|$, $S_{f_2}(\omega)= A_2/|\omega|$, $S_{\fZ }(\omega)= \Az/|\omega|$ and $S_{\fX }(\omega)= \Ax/|\omega|$.  Using this and summarizing the results of \EqRef{eq:c12_cZX} and \EqRef{eq:avf1}, we get
\begin{eqnarray}  \label{eq:ConvCorr}
  A_1  & = & \Az + \Ax/4 - \czx \sqrt{\Az \Ax},  \nonumber \\  
  A_2 & = & \Ax, \nonumber \\ 
  c_{12} &=&  \czx \sqrt{ \frac{ \Az}{A_1}} - \frac{1}{2} \sqrt{ \frac{\Ax}{A_1}}.
\end{eqnarray}
This is Eq. (4) in the main paper.  

\end{document}